\documentclass{elsart}
 \usepackage{graphicx}
\usepackage{amssymb}
\begin{document}
\begin{frontmatter}
\title{How to discriminate easily between Directed-percolation and Manna scaling}
\author{Juan A. Bonachela and Miguel A. Mu\~noz}
\address{Departamento de Electromagnetismo y F{\'\i}sica de la Materia and \\
Instituto de F{\'\i}sica Te{\'o}rica y Computacional Carlos I,\\
Facultad de Ciencias, Universidad de Granada, 18071 Granada, Spain}
\begin{abstract}
Here we compare critical properties of systems in the
directed-percolation (DP) universality class with those of
absorbing-state phase transitions occurring in the presence of a
non-diffusive conserved field, {\it i.e.} transitions in the so-called
Manna or C-DP class. Even if it is clearly established that these
constitute two different universality classes, most of their universal
features (exponents, moment ratios, scaling functions,...) are very
similar, making it difficult to discriminate numerically between
them. Nevertheless, as illustrated here, the two classes behave in a
rather different way upon introducing a physical boundary or
wall. Taking advantage of this, we propose a simple and fast method to
discriminate between these two universality classes. This is
particularly helpful in solving some existing discrepancies in
self-organized critical systems as sandpiles.
\end{abstract}

\begin{keyword}
Self-organization \sep Universality \sep Critical phenomena
\PACS 
05.50.+q \sep 02.50.-r \sep 64.60.Ht \sep 05.70.Ln
\end{keyword}
\end{frontmatter}


The concept of universality in equilibrium phase transitions has been
successfully extended to the broader realm of non-equilibrium, where rather
robust universality classes describing driven diffusive systems, the
roughening of non-equilibrium interfaces, or systems with absorbing states, to
name but a few, are firmly established. For instance, a huge number of models
with absorbing states belong to the directed-percolation (DP) class
\cite{Reviews_AS} which constitutes one of the largest and more robust
non-equilibrium classes.

One specially fascinating family of non-equilibrium critical phenomena
are those appearing in a {\it self-organized} way, {\it i.e.} without
apparent tuning of parameters. Since the introduction of {\it
self-organized criticality} (SOC) \cite{SOC} understanding the
universality (or the lack of it) of ``self-organized'' critical points
became a high priority task. Sandpiles
\cite{SOC,Models}, the archetype example of SOC, are models in which
sandgrains are slowly added to a given site in a lattice, and
redistributed to neighboring sites whenever certain instability
threshold is exceeded.  This may generate cascades of topplings (i.e
{\it avalanches}) which eventually lead to dissipation of grains at
the open boundaries (see \cite{SOC,Models} for model definitions and
reviews). This mechanism drives the system to a stationary, critical
state with power-law distributed avalanches. For our purpose here, it
is important to realize that between successive avalanches sandpiles
are trapped into one out of many possible quiescent or absorbing
states (in which the dynamics is completely frozen). Such a state can
be characterized by the distribution of heights below threshold, which
is a locally conserved and non-diffusive (grains at sites below
threshold do not move) field. After some controversies, it was clearly
established that most (isotropic) stochastic sandpiles \cite{Models}
share the same universal properties \cite{FES,Biham} (the
correspondence is best understood using the {\it fixed-energy
ensemble}; see
\cite{FES}). Indeed, (isotropic) stochastic sandpiles, as well as many
other systems with absorbing states and an additional conserved and
non-diffusive field \cite{Romu}, belong to a unique universality
class, usually called the Manna class \cite{Models} or C-DP class (in
analogy with the nomenclature by Hohenberg and Halperin
\cite{HH}). The essential ingredients (symmetries and conservation
laws) of this class are captured by the following set of Langevin
equations:
\begin{equation}
\label{CDP}  
\begin{array}{rcl}
\partial_t \rho (x,t) & = & a \rho - b \rho^2 + D \nabla^2 \rho + \omega \rho
\, \phi + \sigma \sqrt{\rho}\, \eta(x,t), \\ \partial_t \phi (x,t) & = &
D_{\!\phi} \nabla^2 \rho \mbox{ ,}
\end{array}
\end{equation}
\noindent
where $\rho(x,t)$ is the activity field, $\phi(x,t)$ the background conserved
field, $a, b, D$, $\omega, \sigma$ and $D_{\!\phi}$ constants, and $\eta(x,t)$
a Gaussian white noise \cite{FES,Romu}. This is to be compared with the
standard Langevin equation for generic transitions into absorbing states with
no extra symmetry or conservation law, i.e. the celebrated DP Langevin
equation \cite{Reviews_AS}:
\begin{equation}
\label{DP}  
\begin{array}{rcl}
\partial_t \rho (x,t) & = & a \rho (x,t) - b \rho^2 + D \nabla^2 \rho + \sigma
\sqrt{\rho}\, \eta(x,t),
\end{array}
\end{equation}
\noindent
which differs from C-DP just by the absence of the conserved field. The
previous two equations represent two distinct universality classes as verified
by extensive numerical simulations as well as renormalization group analyses,
and other theoretical considerations \cite{Reviews_AS,FES}.  Given that
Langevin equations are mesoscopic coarse-grained descriptions, in which
irrelevant terms giving corrections to scaling have been excluded, direct
numerical integration of Eq.(\ref{CDP}) and Eq.(\ref{DP}) is an excellent way
to determine critical exponents with good precision \cite{Dornic}.  From such
numerical integrations as well as from extensive numerics performed in
discrete models, the values of critical exponents for these classes can be
obtained (upper rows in Table 1).
\begin{table} [ht!]
\begin{centering}
\begin{tabular}{|l|c|c|c|c|c|}
\hline
&$\eta$&$\delta$&$\tau$&$\tau_{t}$\\
\hline
$DP$&     $0.313(1)$& $0.159(1)$&$1.108(1)$ &$1.159(1)$\\
$C$-$DP$& $0.350(5)$& $0.170(5)$&$1.11(2)$  &$1.17(2)$\\
\hline
\hline
$DP_{abs}$&$0.045(2)$&$0.426(2)$&$1.28(3)$&$1.426(2)$\\
$DP_{ref}$&$0.046(2)$&$0.425(2)$&$1.25(3)$&$1.425(2)$\\
\hline
\hline
$C$-$DP_{abs}$&$-0.33(2)$&$0.85(2)$&$1.56(2)$&$1.81(2)$\\
$C$-$DP_{ref}$&$0.35(3)$ &$0.16(3)$&$1.11(3)$&$1.15(3)$\\
\hline
\end{tabular}
\caption{\footnotesize{Critical exponents for DP and C-DP, without a wall and
in the presence of absorbing and reflecting walls. $\theta$ is the
order-parameter decay exponent: $\langle \rho(t) \rangle \sim
t^{-\theta}$, $\eta$ and $\delta$ are the usual exponents in spreading
experiments (for the growing of the total activity from an initial
seed, $N(t) \sim t^{\eta}$, and the decay of the survival probability,
$P(t) \sim t^{-\delta}$), while $\tau$ and $\tau_t$ are the avalanche
exponents of the size and time distributions respectively
\cite{avalanches}). Values in rows $3$ ($DP_{abs}$) and $4$
($DP_{ref}$) coincide, also those in $2$ (C-DP) and $6$ ($C-DP_{ref}$)
are also equal within errorbars.}}
\label{table_def}
\end{centering}
\end{table}

Note the remarkable similitude between exponents in both classes. The
numerical differences in other critical exponents as $\beta$, $\nu$,
$z$ etc. are not large either. Moreover, other universal features as
scaling functions and moment ratios can also be checked to be very
much alike in these two classes \cite{Lubeck}.  This suggests that
Manna/C-DP exponents could be somehow ``perturbatively'' computable
from DP ones, but unluckily this program has not been completed yet.

Given the general lack of working analytical tools to analyze
microscopic models, in order to assign a given model to one of these
classes one is left mostly at the mercy of numerics which, in the
light of the small differences in exponent values, is not a pleasant
task. Even worse: for some discrete models (in particular, some
sandpiles) corrections to scaling are large and numerics can be
plagued with long transients hiding the true asymptotic behavior. This
is the reason why some original works trying to relate sandpile
criticality with systems with absorbing states concluded, that
exponents were ``compatible'' with DP scaling. Ulterior large scale
simulations revealed systematic differences with DP, and showed rather
unambiguously that sandpiles are generically described by C-DP and its
associated set of exponents, as explained above
\cite{FES,Lubeck}. Nevertheless, there remain some controversial cases in which
it is not easy to distinguish numerically between these two classes
\cite{MDhar} and one has to resort to massive simulations \cite{Jabo1}. In
this paper we propose a way out to this situation: a simple method to
discriminate between these two classes without much computational effort.

The idea is to introduce a wall in the system under scrutiny. It is well known
(already from equilibrium statistical mechanics) that the presence of a
boundary or wall can induce non-trivial ``surface'' critical behavior, which
is also characteristic and specific of each universality class (or
renormalization group fixed point). In the context of systems with absorbing
states, the effect of walls in the DP class has been profusely studied from a
field theoretical perspective \cite{FT,Fro}, using series expansions
\cite{Essam}, density matrix renormalization group \cite{DM}, as well as Monte
Carlo simulations \cite{MCwall}.  As feedback of activity from behind
the wall is impeded (see the nice snapshots in \cite{Fro}) the
structure of the avalanches is strongly affected by the presence of a
wall, be it {\it absorbing} or {\it reflecting}. Owing to this, the
exponents for spreading or avalanches started nearby a wall differ
from the ones without the wall, and are known with good precision
\cite{Fro}. We have simulated a number of (one-dimensional) systems in
the DP class by analyzing the evolution of avalanches (or spreading)
started nearby the wall and measured the exponent values in Table 1
(rows 3 and 4), in excellent agreement with those in the literature
and obeying well-established scaling laws (in particular, the wall
induces the appearance of only one independent new exponent, not
related to standard bulk ones) \cite{Fro}. In accordance with
what is already know \cite{Fro}, all these exponents take the same
values in the DP class for both absorbing and reflecting walls.  Also,
we have studied a model in the DP class with infinitely many absorbing
states, charaterized by a {\it non-conserved} background field
\cite{PCP}. In this case, the critical exponents in the presence of a
wall (either absorbing or reflecting) coincide with those reported
above for the DP class. Even if the background field becomes
inhomogeneous nearby the wall this does not affect critical
properties.

On the other hand, to the best of out knowledge, the effect of walls
in the Manna/C-DP class has not been studied so far. Analogously to
the case before, we have performed Monte Carlo simulations of a family
of different one-dimensional models in this class, including:
stochastic sandpiles (Oslo, Manna, and different variations of them
\cite{Models}), reaction diffusion systems with a conserved non-diffusive
field \cite{Romu}, as well as the Langevin Eq.(\ref{CDP}).

First, we have verified that these types of wall do not alter bulk critical
properties: starting avalanches sufficiently far away from the wall one
recovers the standard (bulk) spreading exponents with as much precision as
wanted. 
\begin{figure}
\includegraphics[height=50mm,width=70mm]{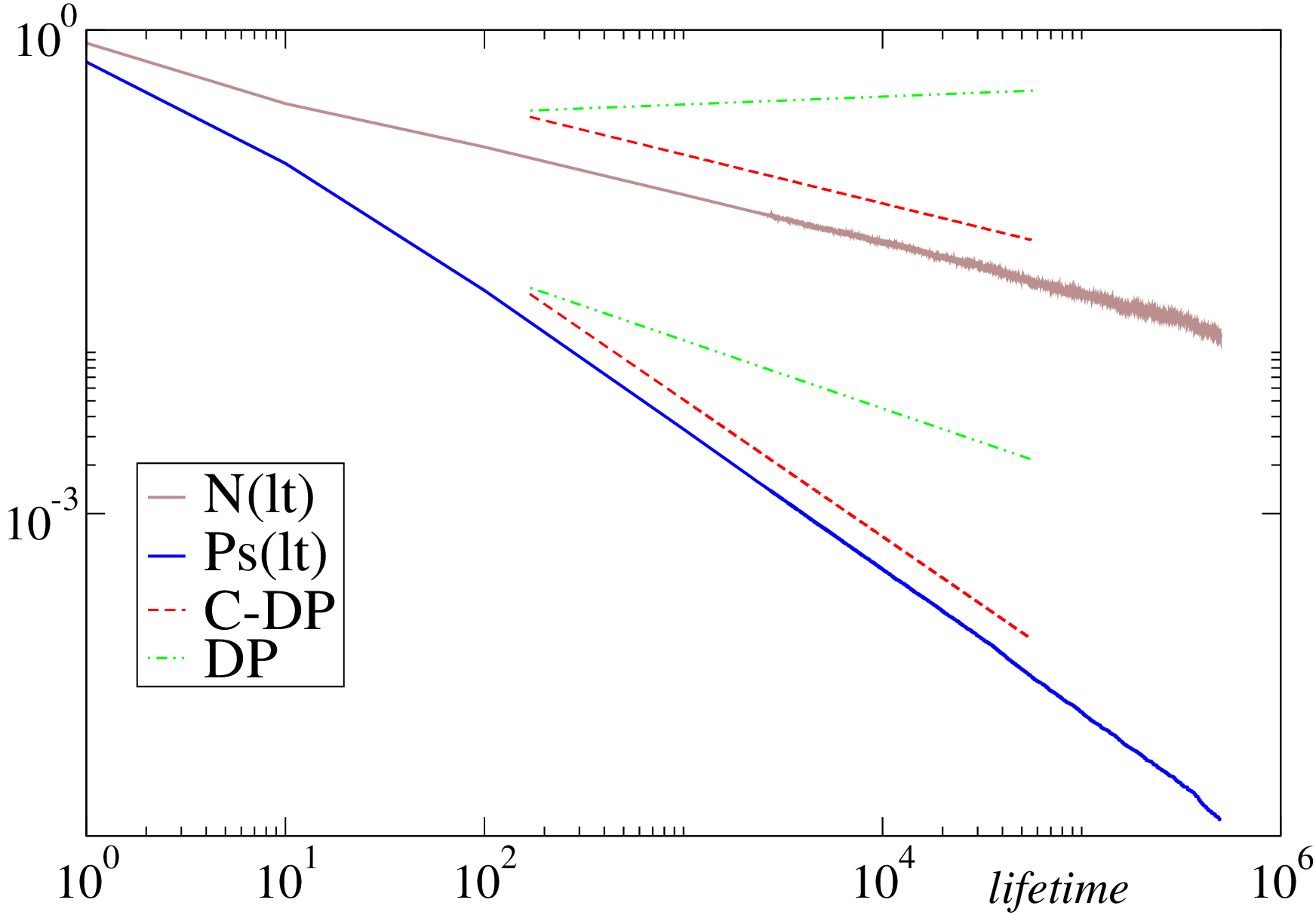}
\includegraphics[height=50mm,width=70mm]{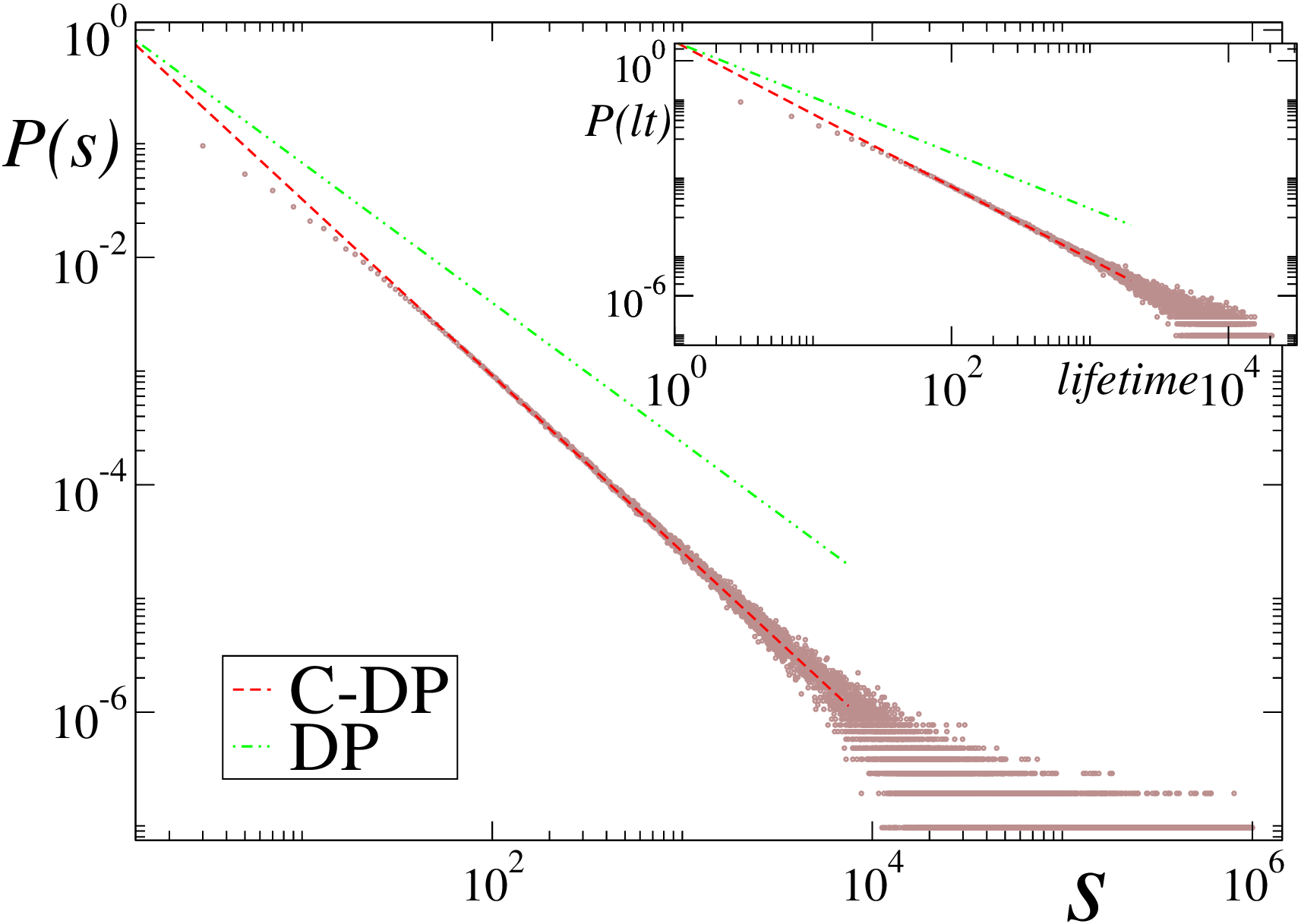}
\caption{Avalanche exponents for the Oslo sandpile model in one dimension,
 averaged over $10^7$ runs (system size $L=2^{15}$) in the presence of an
 absorbing wall.  Left: spreading experiments (see Table 1).  Right: avalanche
 size (main plot) and time (inset) distributions. Green lines mark DP scaling
 while red ones correspond to the best fit, characterizing C-DP/Manna
 surface scaling.}
\end{figure}

Our measurements of exponents (summarized in Table 1; rows $5$ and
$6$) lead to the following conclusions (see Figure 1):
\begin{enumerate}
\item All models whose bulk properties are in the C-DP class, share the same
(universal) surface exponents, which take the values in Table 1. The
corresponding exponents are very robust and universal.

\item Contrarily to the DP class, in this case absorbing and reflecting walls
generate different effects and exponents.
\begin{itemize}
\item Absorbing walls do affect the exponents as happens in DP;
surface exponents differ significatively from their bulk counterparts.
\item Reflecting walls do {\it not} change the exponents with respect to
the ones without a wall; even if feedback of activity from ``behind'' the wall
is also impeded in this case, the background field is enhanced nearby the
reflecting wall, fostering further creation of activity and compensating the
lack of feedback from behind the wall.
\end{itemize}
\end{enumerate}
Therefore, an efficient and simple way to discriminate between DP
and C-DP scaling consists in introducing a wall, either reflecting or
absorbing:
\begin{itemize}
\item If, upon introducing a {\it reflecting wall}, exponents are not changed
with respect to the original ones then the system is C-DP
like. Instead, if they are affected (and take the values in Table 1)
it is DP. The difference between the surface exponents in both classes
is very large (compare, for instance, the values for $\eta$,
$0.045(2)$ versus $0.33(3)$, $700 \%$ larger).

\item In the presence of {\it absorbing walls} the differences in 
the exponent values is also very large. For instance, $\eta =0.045(2)$
for DP while for C-DP the value is $-0.32$ (opposite sign!), and
distinguishing them is a trivial matter.
\end{itemize}

As a straightforward application of these ideas, let us now discuss
the sandpile model introduced by Mohanty and Dhar (MD) in its
self-organized regime \cite{MDhar}. While the directed version of it
is well established to be in the DP class (as happens with other {\it
directed} sandpiles), there has been some controversy about its
behavior in the non-directed case (see
\cite{MDhar} and \cite{Jabo1}). We have simulated the MD sandpile in
the presence of either an absorbing or a reflecting wall. Our main
simulation results are plotted in Figure 2.
\begin{figure}
\includegraphics[height=50mm,width=70mm]{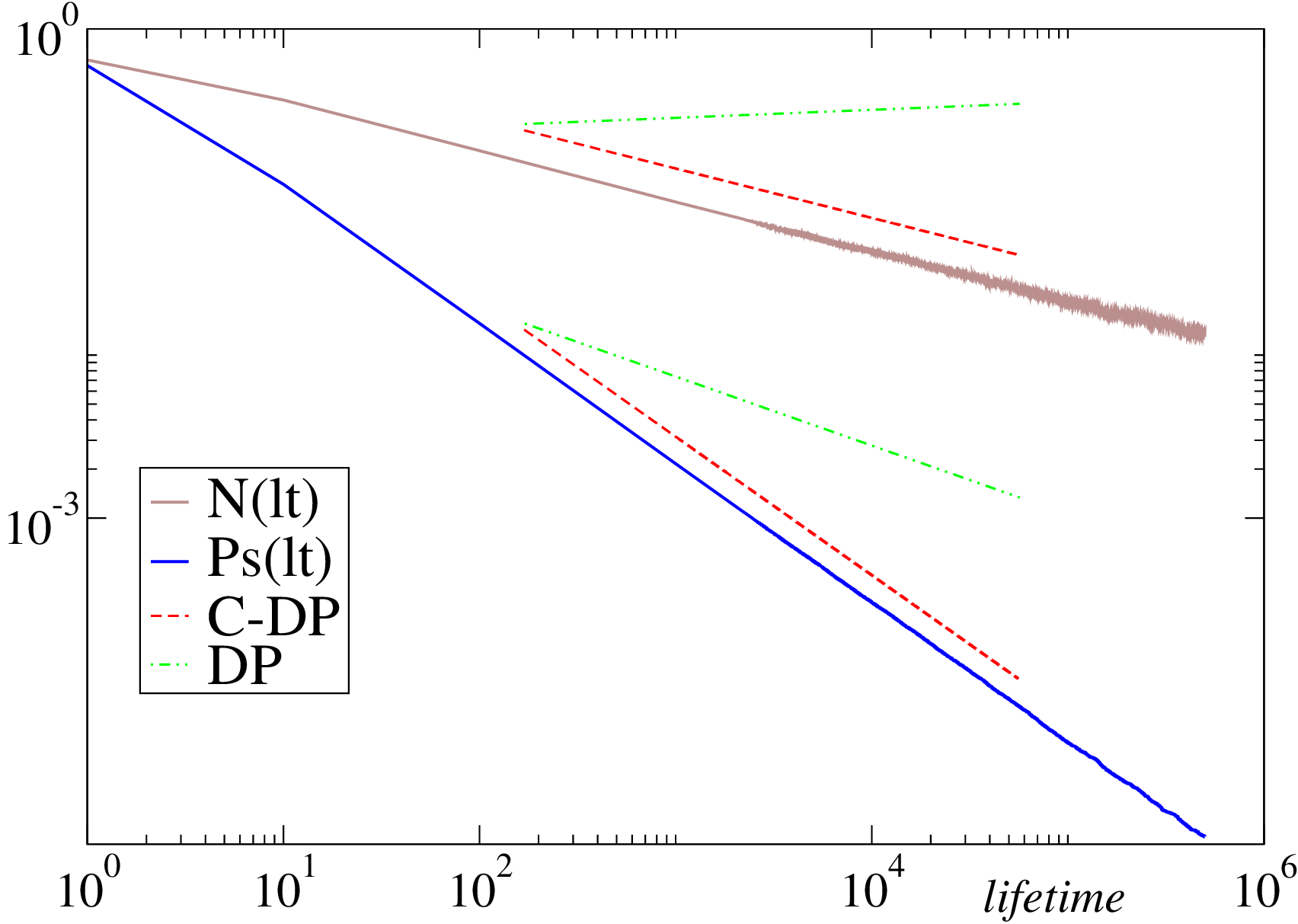}
\includegraphics[height=50mm,width=70mm]{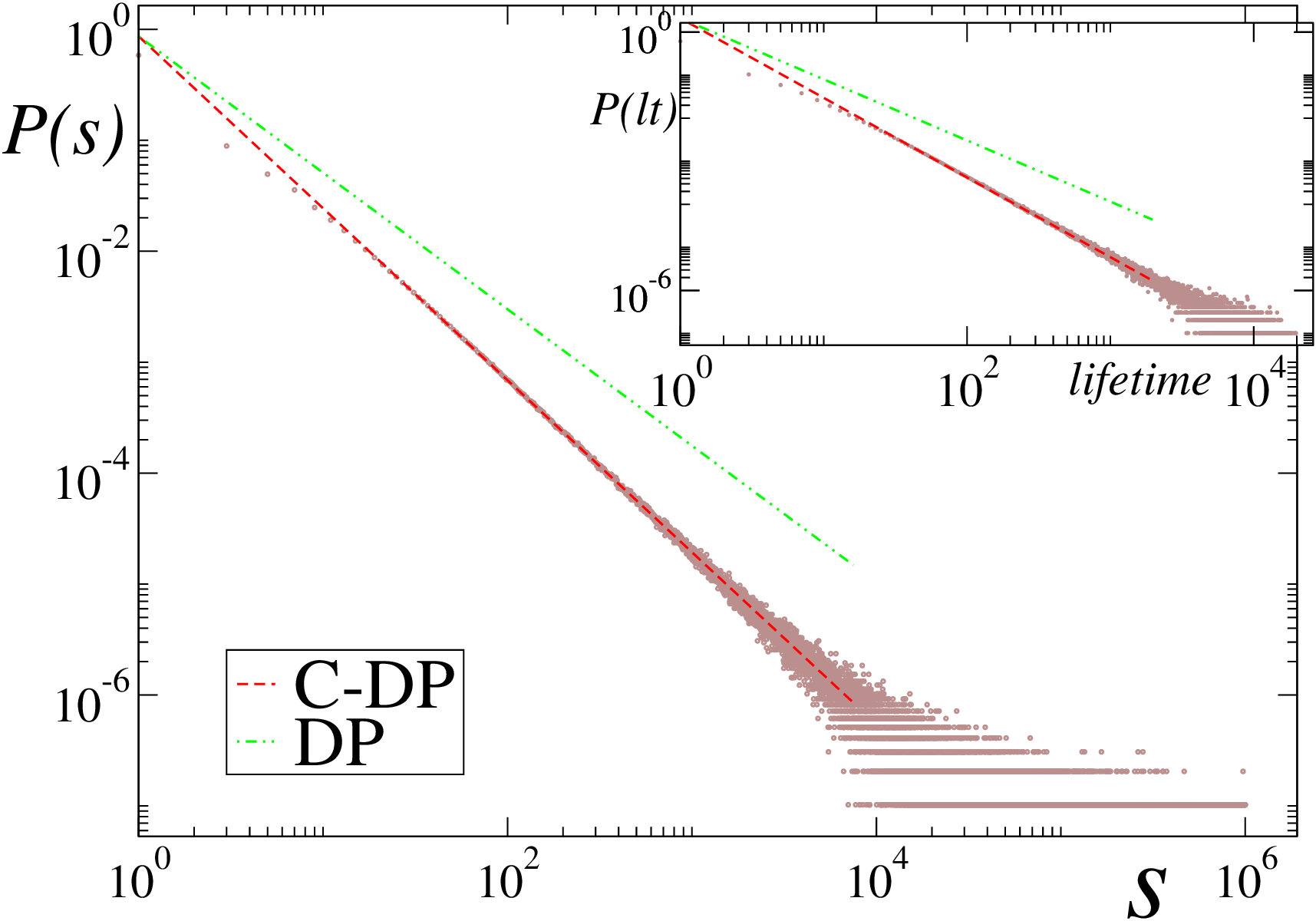}
\caption{Avalanche exponents for the MD sandpile in one dimension, averaged
 over $10^7$ runs (system size $L=2^{15}$) in the presence of an absorbing
 wall.  Left: spreading experiments (see Table 1).  Right: avalanche size
 (main plot) and time (inset) distributions.  The $4$ computed magnitudes are
 in excellent agreement with C-DP values (red lines) and incompatible with DP
 scaling (green lines).}
\end{figure}
First, as before, we have verified that bulk critical exponents are
actually not affected by the presence of a wall: starting avalanches
sufficiently away from the wall one recovers bulk exponents as those
reported in
\cite{Jabo1} with as much precision as wanted. 
Then we determine spreading exponents by starting avalanches in a site
nearby the wall and measuring their corresponding observables as a
function of avalanche time. We also compute the exponents
characterizing avalanche size and time distributions. It is clearly
observed that exponents with and without a reflecting wall coincide
for reflecting walls, while they are different for absorbing walls, in
full agreement with what reported for the Manna/C-DP class. Moreover,
in both cases (i.e. absorbing and reflecting walls) all the computed
exponents coincide with high accuracy with the ones for Manna/C-DP
class in the presence of a wall (see table 1) and exclude DP behavior.

Let us remark that in similar {\it directed models} the introduction of a wall
leads to a ``structured'' background field (not very different from the one
discussed here) that can induce non-trivial results
\cite{Struc1,Struc2}. Indeed, in \cite{Struc1} (non DP) critical exponents
are analytically derived in an exact way by analyzing the structure of
the background and using the directed nature of the process, scaling
laws, and DP exponents. It would be extremely interesting to have an
analogous calculation for the present case, although owing to the lack
of directness this promises to be a much more challenging task.

In summary, by introducing a wall, be it absorbing or reflecting, we have
shown that systems in the, otherwise very similar, DP and Manna/C-DP
universality classes, behave in a very different way. This provides a simple
method to discriminate between these two classes.  In a future work we will
study analytically the effect of walls on systems in the Manna/C-DP class,
complementing the numerics here and providing a more solid theoretical
background to the practical strategy presented in this paper.
\vspace{0.3cm}

We are grateful to D. Dhar, P. K. Mohanty, H. Chat\'e, F. de los
Santos, and I. Dornic for stimulating and illuminating
discussions. Financial support from the Spanish MEyC-FEDER, project
FIS2005-00791 and from Junta de Andaluc{\'\i}a as group FQM-165 is
acknowledged.

\end{document}